\documentclass[aps,pra,twocolumn,superscriptaddress,nofootinbib,showpacs,floatfix,
]{revtex4-1}
\usepackage{bm}
\usepackage{graphicx}
 
\newcommand{\expect}[1]{{\left\langle #1 \right\rangle}}
\newcommand{\expectz}[1]{{\langle #1_z \rangle}}
\newcommand{\dens}[1]{{\rm [#1]}}

\newcommand{\tHe}{\mbox{$^3$He}}

\def \expect#1{{\left \langle #1 \right\rangle}}

\newcommand{\be}{\begin{eqnarray}}
\newcommand{\ee}{\end{eqnarray}}

\begin{document}

 \title{Effects of Nitrogen Quenching Gas on Spin-Exchange Optical Pumping of $^3$He}

\author{B. Lancor and T. G. Walker} 
\affiliation{Department of Physics, University of Wisconsin-Madison, Madison, WI 53706}

\begin{abstract}
We consider the degree of conservation of nuclear spin polarization in the process of optical pumping under typical spin-exchange optical pumping conditions.   Previous analyses have assumed that negligible nuclear spin precession occurs in the brief periods of time the alkali-metal atoms are in the excited state after absorbing photons and before undergoing quenching collisions with nitrogen molecules.  We include excited-state hyperfine interactions, electronic spin relaxation in collisions with He and N$_2$, spontaneous emission, quenching collisions, and a simplified treatment of radiation trapping.
\end{abstract}

\date{\today}


\maketitle

Spin-exchange optical pumping (SEOP)  \cite{WalkerRMP} is a powerful method for spin-polarizing large quantities of $^3$He.  In a typical implementation, rubidium vapor at 180$^\circ$C in 300 cm$^3$ volume glass cells containing several atm of $^3$He and a much smaller partial pressure of molecular nitrogen are optically pumped with intense, circularly polarized  light tuned to the resonance between the 5S$_{1/2}$ and 5P$_{1/2}$  states.  The Rb atoms, spin-polarized to nearly 100\%, collide with the $^3$He atoms and with a small probability per collision the electronic spin-polarization  is transferred to the $^3$He nucleus via a Fermi-contact hyperfine interaction.

The spin-exchange method is slow but efficient, requiring typically ten or more hours to build up \tHe\ polarizations to nearly 80\% \cite{Babcock06}, but in principle requiring absorption of only a few Watts of   optical power.  In practice, however, the laser power demands are found to be substantially higher than this.  It is very important to understand the origin of the inefficiencies in order to obtain the best possible performance for high demand applications such as polarized targets and \tHe\ magnetic-resonance imaging.  For example, we have recently studied the slight breakdown of the atomic angular momentum selection rules due to collisions with He and N$_2$ that  allows absorption of the pumping light by fully polarized atoms \cite{Lancor10,*Lancor10b}.

 An essential but little-studied component of any successful SEOP experiment is 10-100 Torr of nitrogen gas, provided to inhibit relaxation due to radiation trapping \cite{WalkerRMP}.  For applications such as high pressure spin-polarized targets at storage rings \cite{Singh09,*Slifer08}, it is desirable to minimize the nitrogen content of the gas, as it contributes to scattering backgrounds.  In addition,  recent work on the supposedly less demanding application of neutron spin filters has shown unexpectedly large influence of high flux neutron beams on the spin-exchange process  \cite{sharma:083002,Babcock09b}.  Further investigations suggest that the very high observed alkali relaxation rates correlate positively with the nitrogen pressure, again giving motivation to use as little nitrogen as necessary.  It therefore becomes important to quantify the nitrogen density requirements for spin-exchange optical pumping.

This paper presents an analysis of the optical pumping process under conditions typical of spin-exchange optical pumping, in particular considering the effects of excited-state spin-relaxation and hyperfine evolution, and radiation trapping.  The nitrogen gas is a key player in these effects.  In addition to quenching, nitrogen is usually the primary source of fine-structure changing collisions in the excited state, and a contributor to excited-state spin-relaxation \cite{Rotondaro98} and line-broadening \cite{Romalis97}.  All of these effects would be minor, were it not for excited-state hyperfine couplings that cause relaxation of the alkali nuclear spin while the atom is in the excited state.  This effect is usually assumed to be small \cite{WalkerRMP},  but we shall see that, especially for low He pressure applications such as neutron spin-filters, nuclear spin-nonconservation is predicted to become a serious problem when nitrogen pressures are reduced.  Especially when this effect is coupled with relaxation from radiation trapping, we find that the photon demands increase rapidly at low nitrogen pressures.

We begin in Section~\ref{sec:qual} with simple estimates, illustrative simulations, and a discussion of the implications of nuclear spin non-conservation.  We present in Section~\ref{sec:results} the results of the analysis for typical SEOP cell conditions and give predictions for the expected behavior when the nitrogen pressure is reduced to the 10-20 Torr level.   In Section~\ref{sec:quant} we describe our quantitative modelling of the optical pumping process including the effects of excited-state spin-relaxation, fine-structure changing collisions, hyperfine evolution, and  quenching collisions.   Finally, in Section~\ref{sec:radtrap} we consider a simple model of radiation trapping and consider its effects for lower nitrogen pressures.

\section{The effects of Excited-State Spin evolution}

\label{sec:qual}

\subsection{Excited-state spin precession}\label{subsec:precess}
\begin{figure}
\includegraphics[width=3.0 in]{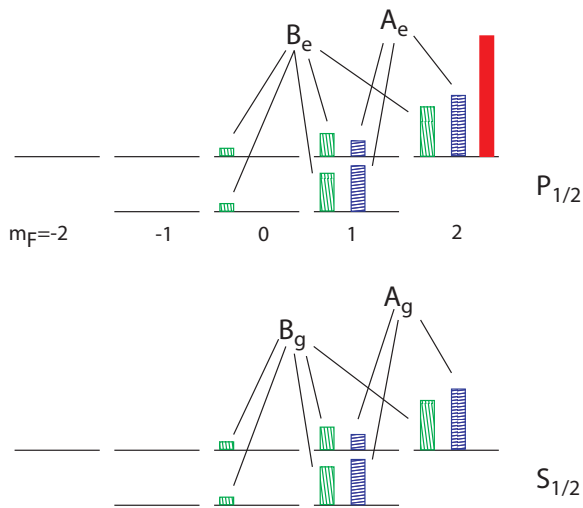}
\caption{If hyperfine precession in the excited state can be neglected, the absorption of a photon from an $m_F=1$ ground state to an $m_F=2$ excited state results in, with fast collisional relaxation in the excited state, a P$_{1/2}$ distribution A$_e$.  Quenching by nitrogen then results in the distribution A$_g$.  Significant hyperfine precession allows more angular momentum to be lost in the excited state, resulting in B$_e$ and B$_g$, thus reducing the optical pumping efficiency.
}\label{fig:elevels}
\end{figure}

  We consider  $^{87}$Rb atoms starting in the $F=2$, $m_F=1$ level of the ground state.  They absorb $\sigma^+$ photons from the laser, promoting them to $F'=2$, $m_F'=2$ in the P$_{1/2}$ excited state, denoted by the solid red bar in Fig.~\ref{fig:elevels}.  Collisions with He atoms   destroy the electron angular momentum within 100 ps at atmospheric pressure, while the nuclear spin is unaffected.  Only states $m_F'=1,2$ containing nuclear spin $3/2$ components can be populated, as shown by the distribution $A_e$ in Fig.~\ref{fig:elevels}.  Subsequent quenching collisions, again conserving nuclear spin, result in population of the fully polarized $m_F=2$ ground state with 50\% probability (distribution $A_g$).
If quenching is not sufficiently rapid, however, hyperfine precession in the excited state begins to transfer nuclear polarization to electronic polarization, thus populating lower $m_F'$ levels, as shown by  distribution $B_e$ in Fig. \ref{fig:elevels}.  Then the probability of reaching the fully polarized $m_F=2$ state is reduced upon quenching collisions.  The key point is that the hyperfine interaction transfers angular momentum from the nucleus to the quickly relaxing electron, reducing the polarization gained per cycle in the optical pumping process.

We can make an estimate of this effect by considering that there is a mean time $\tau$ between  spin-relaxing collisions in the excited state, and the nuclei precess at a rate $A$, where $A {\bf I}\cdot{\bf S}$ is the excited-state hyperfine coupling.  The amount of nuclear spin lost in a single coherence time is given by first-order time-dependent perturbation theory as $\delta I_z\approx (2\pi A \tau)^2I_z$, or the nuclear spin polarization decays with time at a rate $(2\pi A)^2\tau$.  
\begin{figure}[bth]
\includegraphics[width=3.0 in]{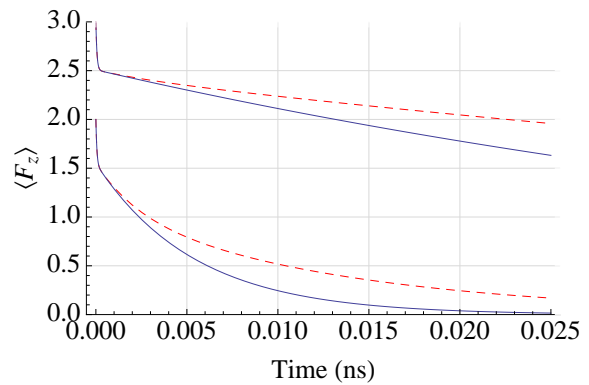}
\caption{Total angular momentum evolution in the excited state, for initially fully polarized P$_{1/2}$ $^{85}$Rb (upper curves) and $^{87}$Rb in 1 amg of He gas and 0.065 amg of N$_2$, assuming no quenching.  Collisions rapidly relax the electronic angular momentum to nearly zero, but hyperfine coupling partially repolarizes the electron so that repeated collisions eventually relax the nucleus as well.  The substantially smaller hyperfine splitting in $^{85}$Rb makes the effect much smaller in that isotope. The dashed curves include fine-structure changing collisions.}\label{fig:excitedevol}
\end{figure}
If the time before a nitrogen quenching collision is $\tau_Q$, there will be $\tau_Q/\tau$ coherent precession intervals while in the excited state.  Thus we expect to lose a fraction   $f_I\approx 1-\exp[(2\pi A)^2\tau\tau_Q]$ of the nuclear spin in the excited state.  At 1 atm He pressure and 50 Torr N$_2$ pressure, the data from Table~\ref{table:multipole} give an estimate $f_I\approx\{0.06,0.53\}$ for $^{85,87}$Rb.  So a significant amount of angular momentum is potentially lost through this effect.  

Figure \ref{fig:excitedevol} shows a more realistic simulation of the total angular momentum as a function of time for atoms that are initially excited to the fully polarized $m'_F=I+1/2$ state.  The calculation includes collisions with  He  and N$_2$, but no N$_2$ quenching collisions or spontaneous emission.  Results are shown with and without collisional transfer to the $P_{3/2}$ state \cite{Rotondaro98}, which slows the nuclear spin-relaxation due to significantly reduced hyperfine interaction in that state.  The electron spin polarization is very rapidly initially lost, making $F_z\approx I$.  But then as the electron and nucleus precess around each other, the rapid He collisions keep removing the electronic angular momentum.  This results in a slow loss of the total angular momentum.  Without N$_2$ quenching to shorten the excited-state lifetime below the 27 ns spontaneous decay lifetime, nearly all the angular momentum would be lost from initially polarized $^{87}$Rb atoms.

 \subsection{Effect of excited-state spin precession on optical pumping}

For optical pumping by monochromatic light tuned to the peak of the D1 resonance, the rate at which the atoms absorb photons is very nearly  $R(1-2\expect{S_z})=R(1-P)$, where the optical pumping rate $R$ is the rate at which unpolarized atoms absorb light \cite{WalkerRMP,Lancor10}.  Absorption of circularly polarized photons increases the total angular momentum per atom from $\hbar\expectz{F}$ to  $\hbar(\expectz{F}+1)$.  The excited-state collisional relaxation of the electronic angular momenta  removes $\hbar/2$, and the nuclei lose  $f_I\expect{I_z}\hbar$ due to excited-state hyperfine precession.  After quenching,  the angular momentum of the atoms is therefore $\expect{{ F}_z}'=\expect{{F}_z}+1/2-f_I\expect{I_z}$. The optical pumping process therefore obeys the rate equation
\begin{eqnarray}
{d\expect{F_z}\over dt}&=&R\left(1-P\right)\left({1/ 2}-f_I\expect{I_z}\right)\\
&=&{R\over 2}\left(1-P\right)\left(1-f_I\epsilon P\right)
\label{rateeqn}\end{eqnarray}
where $\epsilon=\expect{I_z}/\expect{S_z}$ varies between $4I(I+1)/3$ at low polarizations to $2I$ at high polarizations for atoms in spin-temperature equilibrium \cite{WalkerRMP}.  (For a natural mixture of the two Rb isotopes, it ranges from 9.8 at low polarizations to 4.44 at high polarizations.)
At high polarizations, $P\sim 1$, the excited-state nuclear spin relaxation is effectively a reduction in the optical pumping rate from $R$ to $R(1-f_I\epsilon)$.  

\begin{figure}[htb]
\includegraphics[width=3.0 in]{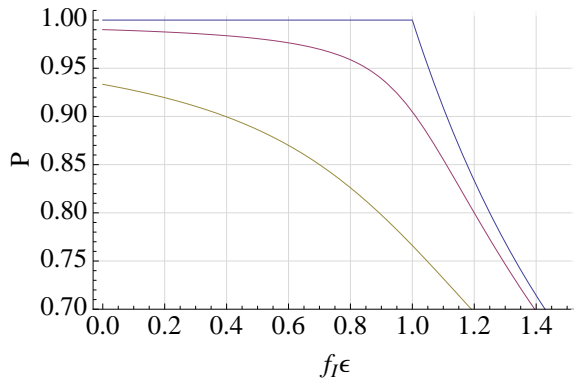}
\caption{The steady-state polarization as a function of the product $f_I\epsilon$, which is approximately twice the angular momentum lost by the nucleus during the excited-state evolution. The three curves are, top to bottom,  for $R=\infty$, $R=100\Gamma_{\rm sd}$, and $R=14\Gamma_{\rm sd}$.}
\label{fig:maxpol}
\end{figure}

An interesting and important feature of Eq.~\ref{rateeqn} is that if $f_I\epsilon>1$, the atom cannot be fully spin-polarized even at infinitely large pumping rates.  The maximum steady-state polarization is   $P_{\max}=\min(1,1/f_I\epsilon)$, shown in Fig.~\ref{fig:maxpol}.

\begin{figure}[htb]
\includegraphics[width=3.0 in]{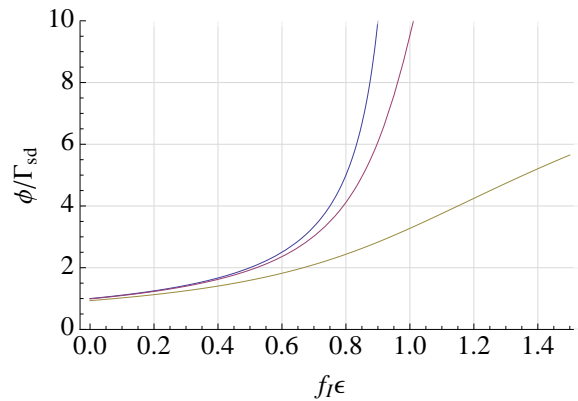}
\caption{The photon demand as a function of the product $f_I\epsilon$,  normalized to the ground-state spin-relaxation rate $\Gamma_{\rm sd}$,   the value that would be obtained in the absence of excited-state nuclear spin-precession. The three curves, top to bottom,  are for $R=\infty$, $R=100\Gamma_{\rm sd}$, and $R=14\Gamma_{\rm sd}$.}
\label{fig:scattrate}
\end{figure}

The steady-state photon absorption rate is, not too surprisingly, also increased by the excited-state relaxation.  Adding a ground-state spin-relaxation term $-\Gamma_{\rm sd}\expect{S_z}$ to Eq.~\ref{rateeqn}, we can express the absorption rate or photon demand as
\be
\phi=R(1-P_{})={\Gamma_{\rm sd}P_{}\over 1-f_I\epsilon P_{}}
\label{absrate}\ee
Again, as $f_I\epsilon P$ approaches 1, the absorption rate increases dramatically, as shown in Fig.~\ref{fig:scattrate}.

\section{Results}\label{sec:results}

\begin{figure}[htb]
a)\hfill $\;$

\includegraphics[width=3.0 in]{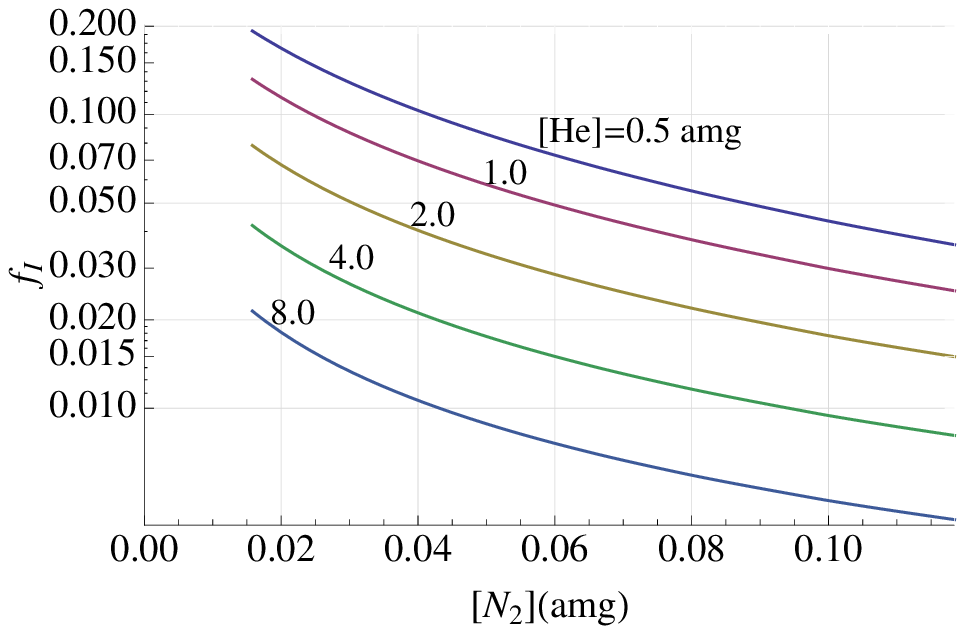}

b)\hfill $\;$

\includegraphics[width=3.0 in]{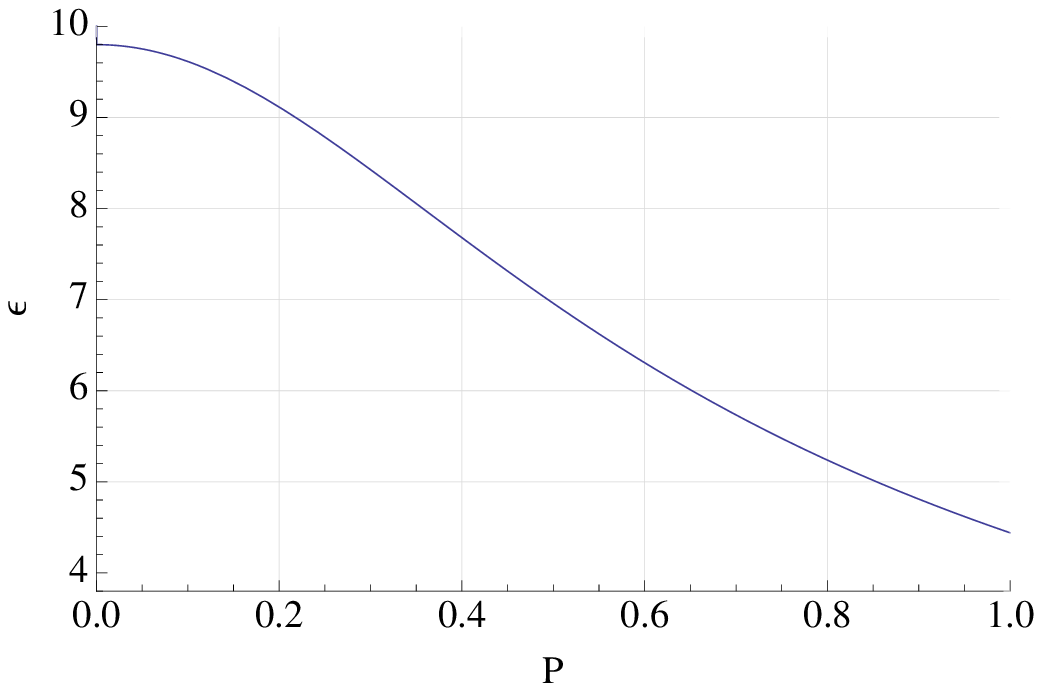}

\caption{a) Calculated fraction of nuclear spin lost, $f_I$, in the excited state during the optical pumping cycle, at various nitrogen and He densities. b) Paramagnetic coefficient  $\epsilon$ for natural abundance Rb vapor in spin-temperature equilibrium, as a function of electron spin-polarization.}
\label{fig:fIresults}
\end{figure}

We have used the optical pumping model described in the next section to evaluate $f_I$ for natural Rb vapor, with 72\% $^{85}$Rb and 28\% $^{87}$Rb, for ranges of He and N$_2$ pressures of interest for SEOP. The results are shown in Fig.~\ref{fig:fIresults}, along with $\epsilon(P)$.

For 8 amg He cells, even at low N$_2$ density the product $f_I\epsilon<0.1$, so the excited-state relaxation has small effects on the polarization and scattering rates.  However, at 1 amg the effects are significant; $f_I\epsilon$ ranges from $0.14$ to 0.63 over the range of nitrogen densities considered.

Figure~5 can be used to easily estimate the values of $f_I$ and $\epsilon$ for a wide range of \dens{He} and \dens{N$_2$}.  The values of $\Gamma_{\rm SD}$  can be estimated from Ref.~\cite{Chen07}.  Then Eqs.~\ref{rateeqn} and \ref{absrate} can be used to calculate  the alkali polarization and photon demand.

\begin{figure}[htb]
\includegraphics[width=3.0 in]{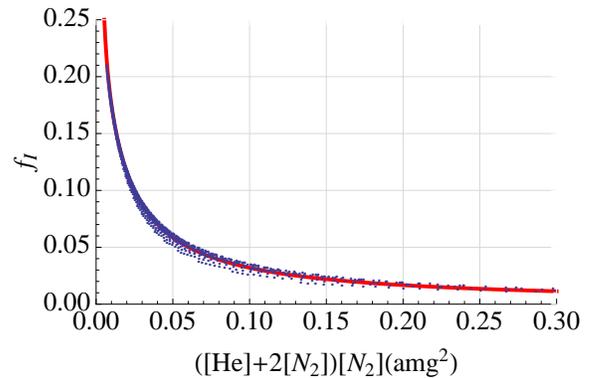}
\caption{The calculated fraction of nuclear spin lost in the excited state as a function of the product of the quenching rate (proportional to \dens{N$_2$}) and the spin-relaxation rate (proportional to $2\dens{N_2}+\dens{He}$). The calculations are well fit by the curve $f_I= 1-e^{-0.00138/x}+(0.00213/x) e^{-0.0152129/x}$ where $x=\dens{N_2}(2\dens{N_2}+\dens{He})$ in units of amg$^2$.}
\label{fig:universal}
\end{figure}

According to the arguments in Sec.~\ref{subsec:precess},  the fraction of nuclear spin lost in the excited state should be a function of the product of the quenching rate and the spin-relaxation rate.  In Fig.~\ref{fig:universal} we show the values of $f_I$ from Fig.~\ref{fig:fIresults} as a function of the product $\dens{N_2}(\dens{He}+2\dens{N_2})$ at a discrete set of points that span the full range of nitrogen and helium densities.  The factor of 2 is chosen to reflect the approximate doubling of N$_2$ relaxation rates as compared to He.  Indeed, when presented this way the results nearly collapse to a single curve.

\section{Quantitative Optical Pumping Model}\label{sec:quant}

We have developed a detailed model of the optical pumping portion of the spin-exchange process using the formalism of Ref.~ \cite{HJW}, hereafter cited as HJW.  The processes of optical pumping and collisions are represented by matrices in Liouville space, and the density matrix is represented as a vector.  See also Ref. \cite{Appelt98} for a similar approach.

\subsection{Atom-light interaction} We assume circularly polarized optical pumping light of a Gaussian spectral profile and with a bandwidth of 100 GHz, a value now common to many SEOP experiments  \cite{Chen07}.  The full hyperfine structure of the ground 5S$_{1/2}$ and excited 5P$_J$ states are taken into account, and a magnetic field of 5 Gauss is assumed to be applied along the laser propagation direction.  Pressure broadening of the resonance lines is taken to be Gaussian, with widths from Ref.~\cite{Romalis97}.  The  pumping rate $R$  is assumed to be an adjustable parameter.  With the large laser linewidth as compared to the hyperfine structure, $R$ is nearly the same for the two Rb isotopes.  

\subsection{Excited state evolution}  The excited-state density matrix $\rho^e$ is sourced by optical excitation, evolves due to hyperfine interactions and collisions with He and nitrogen atoms (rate $\Gamma_c$), and decays via nitrogen quenching and spontaneous emission at rates $\Gamma_Q$ and $\Gamma_s$:
\be 
\dot\rho^e=RA_p^{eg}\rho^g-(iH^{e\tiny\copyright}+\Gamma_cA^{ee})\rho^e-(\Gamma_{\rm s}+\Gamma_Q)\rho^e
\label{excitedrho}
\ee
Assuming no coherence in the ground-state density matrix $\rho^g$, the excitation matrix $A_{\rm p}^{eg}$, HJW (6.35), couples only to excited-state populations.  The matrix $H^{e\tiny\copyright}$, HJW (5.90), is equivalent to a commutator in Schr\"odinger space.  It is diagonal in the Liouville representation and contains the Bohr frequencies associated with the excited-state hyperfine structure.   Due to fine-structure changing collisions, included in $A^{ee}$, both P$_{1/2}$ and P$_{3/2}$ evolution are important.  The matrix $A^{ee}$ and the collision rate $\Gamma_c$ will be discussed further in Sec.~\ref{sec:coll}.  

It is convenient to combine the last two terms in Eq.~\ref{excitedrho} into an excited-state evolution matrix $G^{ee}=iH^{e\tiny\copyright}+\Gamma_cA^{ee}+\Gamma_{\rm s}+\Gamma_Q$.  Then the steady-state solution to   Eq.~\ref{excitedrho} is
\be 
\rho^e=R[G^{ee}]^{-1}A_{\rm p}^{eg}\rho^g
\ee

\subsection{Excited-state spin relaxation}\label{sec:coll}  Collisions with nitrogen and He cause excited-state spin-relaxation, including transfer between fine-structure levels.  These collisions are assumed to be binary and of sufficiently short duration to  conserve nuclear spin.  The relaxation is conveniently described by a multipole expansion of $\rho^e$ \cite{Baylis79}, or equivalently by expanding $A^{ee}$ in terms of multipole projection operators $\Pi_l^{J'J}$  of HJW Sect. 11.1:
\be 
A^{ee}=\sum_{lJJ'}\alpha_l^{J'J}\Pi_l^{J'J}
\ee 
The expansion coefficients can be deduced from the state-to-state collision experiment of Rotondaro and Perram  \cite{Rotondaro98}.  The coefficients with $J'\ne J$, $l=0,1$ describe fine-structure changing collisions.  The coefficients with $J'=J=3/2$, $l=1,2,3$, describe spin relaxation in the P$_{3/2}$ state.  The coefficient $\alpha_1^{1/2,1/2}$ represents spin relaxation in the P$_{1/2}$ state. The multipole relaxation rates $\Gamma_c\alpha_l^{J'J}$, chosen to reproduce the results of Ref.~ \cite{Rotondaro98}, can be calculated from Table~\ref{table:multipole}. 

While the fine-structure-changing and multipole relaxation processes conserve nuclear spin, they produce coherences between the different excited-state hyperfine levels.  The subsequent precession of these hyperfine coherences results in loss of angular momentum from the Rb nuclei.  To illustrate this effect, we show in Fig.~\ref{fig:excitedevol} the nuclear and electronic angular momenta as a function of time for an atom initially excited to the P$_{1/2}$ $F'=2$ state of $^{87}$Rb, assuming no quenching.  It is clear that, even in the absence of radiation trapping effects, quenching is necessary to preserve the nuclear angular momentum in the excited state.

\subsection{Optical Pumping} 

We now consider how optical pumping of the ground-state is affected by the excited-state spin-precession.    There are two contributions to the optical pumping.  Depopulation pumping removes atoms from the ground state, while repopulation pumping replenishes the ground state from the excited-state  either by quenching or spontaneous emission.

  This repopulation pumping obeys
\be 
\dot\rho^g_{\rm RP}=\left(\Gamma_s A_s^{ge}+\Gamma_QA_Q^{ge}\right)\rho^e=G^{ge}\rho^e
\ee
The spontaneous emission matrix $A_s^{ge}$ is given by HJW (5.50), while the quenching matrix is
\be
A_Q^{ge}=\sum_{J} \Pi_0^{SJ}
\ee
We are assuming that the quenching process fully transfers nuclear polarization from the excited state to the ground state, with no transfer of electronic polarization.  The quenching rates can be determined from Table~\ref{table:multipole}.

The net evoluation from optical pumping  is the sum of the depopulation and repopulation pumping terms.   Using HJW 6.16-18 for the repopulation pumping, we have
\begin{eqnarray}
\dot\rho^g_{\rm OP}&=&-RA_{\rm p}^{gg}\rho^{g}+G^{ge}\rho^{e}\\
&=&-RA_{\rm p}^{gg}\rho^{g}+RG^{ge}[G^{ee}]^{-1}A_{\rm p}^{eg}\rho^g
\\
&=&-RA_{\rm OP}\rho^{g}
\label{groundrho}
\end{eqnarray}
Under typical pumping conditions, no Zeeman or hyperfine coherences are generated, so we assume that the ground state density matrix is well represented by populations alone, so that  evolution due to light shifts and ground-state hyperfine interactions are not necessary to include in Eq.~\ref{groundrho}.

\subsection{Ground-state spin-randomization}
There are a variety of important ground-state spin-relaxation mechanisms at work in spin-exchange optical pumping.  The most important, spin-exchange collisions between the alkali-metal atoms, conserves the total angular momentum and produces a spin-temperature distribution.  These will be treated in Sec.~\ref{RbRb}.

Depending on conditions, the most important relaxation mechanisms that do not conserve the total angular momentum are typically electron randomization due to the spin-rotation interaction in Rb-He  and Rb-N$_2$ collisions \cite{Baranga98c}, electron randomization due to the spin-axis interaction in Rb-Rb collisions \cite{Kadlecek01}, and the formation of Rb$_2$ molecules \cite{Kadlecek98,Erickson00}.  As the focus of this paper is not on these mechanisms, we lump them together into an effective electron randomization rate $\Gamma_{\rm sd}$ and simply represent the ground-state relaxation as
\be 
\dot\rho^g_{\rm SR}=-\Gamma_{\rm sd}A^{gg}_{\rm sd}\rho^g
\label{relax}
\ee
where the spin-damping matrix $A^{gg}_{\rm sd}$ is given in HJW 6.88.
Combining the optical pumping and spin randomization gives 
\be 
\dot\rho^g&=&-G^{gg}\rho^g\\
G^{gg}&=&RA_{{\rm OP}}+\Gamma_{\rm sd}A^{gg}_{{\rm sd}}
\ee
for the ground-state density matrix evolution.

\label{spintempevol2}
\begin{table}
\begin{tabular}{l|r|r|r|r|}
Multipole process (5P$_J\rightarrow$5P$_{J'}$)&l=0&1&2&3\\\hline
5P$_{3/2}$+He$\rightarrow$5P$_{3/2}$+He&&240&280&200\\
5P$_{3/2}$+N$_2\rightarrow$5P$_{3/2}$+N$_2$&&214&283&266\\
5P$_{1/2}$+He$\rightarrow$5P$_{1/2}$+He&&32&&\\
5P$_{1/2}$+N$_2\rightarrow$5P$_{1/2}$+N$_2$&&65&&\\
5P$_{1/2}$+He$\rightarrow$5P$_{3/2}$+He&0.072&0*&&\\
5P$_{1/2}$+N$_2\rightarrow$5P$_{3/2}$+N$_2$&10.1&-0.3&&\\ \hline \multicolumn{5}{c}{}\\
\end{tabular}
\begin{flushleft}\hspace*{0.2 in}
\begin{tabular}{l|r|}
Quenching&$\sigma_Q$\\ \hline
5P$_{3/2}$+N$_2\rightarrow$5S$_{1/2}$+N$_2$&43 \\
5P$_{1/2}$+N$_2\rightarrow$5S$_{1/2}$+N$_2$&58 \\ 
\end{tabular}\end{flushleft}
\caption{(top) Multipole relaxation cross sections $\sigma_l^{J'J} $(in \AA$^2$), adapted from Ref.~ \cite{Rotondaro98}.  The asterisk denotes an assumed  quantity.  The multipole relaxation rates for a gas of density $\dens{G}$ are $\Gamma_c\alpha_l^{J'J}=\sigma_l^{J'J}v\dens{G}$, where $v=\sqrt{8kT/\pi\mu}$ is the mean thermal velocity for atom pairs of reduced mass $\mu$ at temperature $T$.  (bottom) Quenching rates $\Gamma_Q=\sigma_Qv\dens{N_2}$.}\label{table:multipole}
\end{table}

\subsection{Rb-Rb spin-exchange}\label{RbRb}
Spin-exchange collisions between Rb atoms conserve the total angular momentum, but redistribute the spin and nuclear Zeeman populations toward a spin-temperature distribution \cite{Happer72,HJW}
\be 
\rho_{\rm ST}(P)=Z(P)^{-1}e^{-\beta(P) F_z}
\ee
where the spin-temperature parameter $\beta$ is determined by the Rb electron spin polarization $P$ via $P=\tanh(\beta/2)$, and $Z$ is a normalizing factor.
  At the high Rb densities used in spin-exchange experiments, the Rb-Rb spin-exchange rates  dominate any of the other rates in the system and so the ground-state density matrix should be well described by a spin-temperature distribution.

 To this point, the two isotopes were treated separately.  The rapid spin-exchange collisions directly couple the two isotopes, so their density matrices are not independent; the electron spin-polarizations are equal.  In the spin-temperature limit, the simplest way to treat the spin-exchange effects is to consider the isotopic fraction weighted total angular momentum
\be 
\expect{{ F}_z}=\sum_i \eta_i\expect{F_{zi}}\label{isoavg}
\ee
where $\eta_i$ is the isotopic abundance of isotope $i$.   (In the following, analogous isotope subscripts will be added to various quantities as needed.)  Then the rate equation for $\expect{{F}_z}$ is 
\be
\dot\expect{{F}_z}=-\sum_i\eta_i{\rm Tr}[F_{zi}G^{gg}_i\rho^{g}_{i,\rm ST}(P)]
\label{spintempevol}
\ee
Eq. \ref{spintempevol} is a non-linear equation, as $\expect{{ F}_z}$ and $P$ are non-linearly related, but it is easy to find the steady state solution by varying $P$ until Eq.~\ref{spintempevol} is zero.

Having found $P$ and the absorption rate $\phi=R {\rm Tr}({A_{\rm p}^{gg}\rho^g)}$, we calculate $\epsilon$  and rearrange Eq.~\ref{absrate} to get
\be
f_I={R-(R+\Gamma_{\rm sd})P\over \epsilon\phi}
\ee
where $R$ is the isotopically averaged pumping rate. 

The parameterization $f_I$ is only useful if it relatively insensitive to the polarization.  Indeed, we find that it generally decreases by 10\% or less as the polarization is decreased (by decreasing the pumping rate in the model) to values well below 50\%.  The results in Fig.~\ref{fig:fIresults} were calculated at a pumping rate of $R=14\Gamma_{\rm sd}$.

\section{Radiation Trapping}\label{sec:radtrap}

 Due to the extreme optical thickness of SEOP cells, often 100 optical depths, any resonance light emitted in the optical pumping process is re-absorbed before leaving the cell.  Since this light is nearly unpolarized, it acts as an efficient relaxation agent with the consequence that optically thick cells cannot be polarized without some means of counteracting the relaxation from radiation trapping.  One solution, proposed by Peterson and Anderson \cite{Peterson91}, is to apply a large magnetic field so that the fluorescent light has an emission profile with at least one component that is non-resonant with spin-polarized atoms.  This method is very effective for optical pumping of dense, non-pressure-broadened cells and has been in intensive use for decades at polarized ion sources around the world  \cite{Anderson79,Levy98}.  For spin-exchange optical pumping, the required magnetic fields  would be on the order of 1 Tesla, which is impractical.

The second means of circumventing radiation trapping is to collisionally quench any optically excited atoms in a time short compared to the spontaneous lifetime of the excited state \cite{Happer72}.  Nitrogen is by far the most convenient molecule for this purpose.  It is one of the few molecules that does not chemically react with hot alkali vapor, and it has vibrational excitations that are nearly resonant with the 1.5 eV alkali resonance lines, resulting in very large ($\sim 50$ \AA$^2$) quenching cross sections.  With tens of Torr of nitrogen pressure, the probability of spontaneous emission can be reduced by a factor of ten or more, allowing the vapor to become optically pumped with only a minor impact on the laser power demand from radiation trapping. 

  To our knowledge, radiation trapping has not been explicitly treated in models of spin-exchange optical pumping.  In general, radiation transport is a highly nonlinear problem.  We propose the following simplified model to account for it.

Since the line-center optical thickness of most SEOP cells is on the order of 100, to a good approximation photons emitted in the line core will be reabsorbed before leaving the cell.  These photons, which are essentially unpolarized, will therefore be absorbed by nearby atoms and they will act as an additional spin-relaxation mechanism.  In the limit that the quenching is rapid compared to spontaneous decay, the probability of the absorption of a photon from the optical pumping laser resulting in more than one re-emitted photon is very small.  Thus since photons are being spontaneously emitted at a rate ${\cal R}={\rm Tr}[\Gamma_s\rho^e$], on average the absorption rate of the unpolarized emission photons must be the same.

We therefore approximately model the effects of radiation trapping by 
\be 
\dot\rho^g_{\rm RT}=-{\cal R}(1-G^{ge}[G^{ee}]^{-1}A^{eg}_{\rm RT})\rho^g=-{\cal R}G_{\rm RT}\rho^g
\label{radtrap}
\ee
where the matrix $A^{eg}_{\rm RT}$ is generated in the same manner as the spontaneous decay matrix $A_{\rm s}^{ge}$, only with the roles of excited and ground states reversed.  Thus in the notation of HJW
\be
A_{\rm RT}^{eg}={2\over 3}\sum_{J j}f_J\Delta^T_j\otimes\Delta^\dag_j
\ee 
where $f_J$ is the fraction of light emitted by atoms in the excited state with angular momentum $J$.  

Thus we find the steady-state solution for our full optical pumping model by solving
\be 
0=\sum_i\eta_i{\rm Tr}[F_z(G^{gg}_i+{\cal R}G_{i,\rm RT})\rho^{g}_{i,\rm ST}(P)]
\ee
Since ${\cal R}$ depends on $P$, it is necessary to iterate a few times to obtain a consistent solution.

A simple representation of the effects of radiation trapping in the spirit of Eq.~\ref{rateeqn} is given by the following argument.  Upon absorption of an unpolarized photon from another atom, the subsequent excited-state evoluation causes virtually all the electronic angular momentum and a fraction $f_I\expect{I_z}$ of the nuclear angular momentum to be lost.  Thus the  total change in angular momentum due to absorption of a radiated photon is $\expect{S_z}+f_I\expect{I_z}=(1+f_I\epsilon)P/2$.  The radiated photons are emitted at the rate $R(1-P)\Gamma_s/(\Gamma_s+\Gamma_Q)$ and must therefore be absorbed by other atoms at that same rate.  In terms of the spontaneous emission probability $f_s={\Gamma_s/( \Gamma_s+\Gamma_Q)}$, the optical pumping rate equation then becomes
\be
{d\expect{F_z}\over dt}={R\over 2}\left(1-P\right)\left(1-P\left[f_I' \epsilon+f_s\right]\right)\label{rate2}
\ee
where
\be
f_I'=f_I\left(1+f_s\right).
\ee
  The photon demand of Eq.~\ref{absrate} becomes
\be
\phi=R(1-P_{})={\Gamma_{\rm sd}P_{}\over 1-\left[f_I' \epsilon+f_s\right]P}
\label{absrate2}\ee
The two terms in square brackets in Eqs.~\ref{rate2} and \ref{absrate2} correspond to the nuclear and electron spin lost in the excited state.

\section{Discussion and Conclusions}

\begin{figure}[tb]
\includegraphics{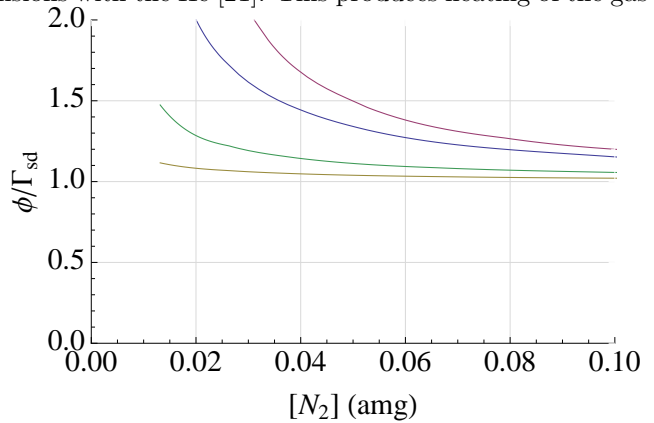}
\caption{The  photon demand normalized to its ideal value.  The upper two curves are for 1 amg He density, the lower two for 8 amg.  The upper curve of each pair includes both radiation trapping and excited-state relaxation effects, while the lower includes only excited-state relaxation.}\label{fig:PhotonDemand}
\end{figure}

A convenient figure of merit for evaluating these effects is the ratio of the photon demand $\phi$ to the ground-state relaxation rate $\Gamma_{\rm sd}$; at high Rb polarization this ratio would be unity in the absence of  excited-state nuclear relaxation and radiation trapping.  Fig.~\ref{fig:PhotonDemand} shows the calculated  photon demand as a function of N$_2$ density, with and without radiation trapping, for two different representative He densities.

At high pressures we see that radiation trapping is somewhat more important than excited-state hyperfine precession, but the total effect is less than 10\% with 0.05 amg or more of nitrogen.  At 1 amg densities, however, the excess photon demands can become quite serious if insufficient nitrogen is present.  The photon demands doubles below 0.03 amg and is still about 20\% larger than the ideal value at 0.1 amg.

While we have concentrated in this paper on the importance of nitrogen for suppressing excess relaxation from hyperfine precession and radiation trapping, there are at least four other important considerations for spin-exchange optical pumping that concern the nitrogen density and we mention them here for completeness.  First,  the energy stored in the vibrational degrees of freedom of the N$_2$ following quenching collisions is dissipated by collisions with the He \cite{Walter01}.  This produces heating of the gas to temperatures that may exceed that of the wall by more than 100$^\circ$C.  Since Rb-He spin-relaxation is strongly temperature dependent \cite{Baranga98c}, this effect should be taken into account when considering the actual photon demand.  

The second additional effect is spin-relaxation in ground-state Rb-N$_2$ collisions which, though usually a small contributor to $\Gamma_{\rm sd}$,  come into play if the N$_2$ fraction becomes too large \cite{Baranga98c,Chen07}.  Generally one wishes to work at N$_2$ densities small enough that the N$_2$ contribution to spin-relaxation is a small effect.  Third, we note that N$_2$ also is a contributor to pressure broadening. 

Finally, we have recently demonstrated that He and N$_2$ collisions allow fully polarized Rb atoms to absorb resonant circularly polarized D1 light \cite{Lancor10,*Lancor10b}, an effect forbidden for free atoms.  This effect also has important consequences for the photon budget in spin-exchange optical pumping.  It is our intention to combine all these effects together in the near future for an analysis of the photon budget for spin-exchange optical pumping.

\acknowledgements{
This work was supported by the U.S. Department of Energy, Office of BasicÊ Energy Sciences, Division of Materials Sciences and Engineering, underÊ award DE-FG02-03ER46093
}

\bibliography{/Users/Thad_Walker/Research/thadbibtex/spinexchange}
\end{document}